\documentstyle[aps,pre,twocolumn,epsfig,floats]{revtex}

\newcommand{\beq}{\begin{equation}}
\newcommand{\eeq}{\end{equation}}
\newcommand{\bdis}{\begin{displaymath}}
\newcommand{\edis}{\end{displaymath}}
\newcommand{\bea}{\begin{eqnarray}}
\newcommand{\eea}{\end{eqnarray}}
\newcommand{\barr}{\begin{array}}
\newcommand{\earr}{\end{array}}

\newcommand{\equ}[1]{(\protect\ref{#1})}

\begin{document}
\draft 
\wideabs{
\title{Corrections to scaling in the forest-fire model}
 
\author{Romualdo Pastor-Satorras and Alessandro Vespignani}

\address{The Abdus Salam International Centre 
  for Theoretical Physics (ICTP)\\
  Condensed Matter Section\\
  P.O. Box 586, 34100 Trieste, Italy\\
  \date{\today}}

\maketitle

\begin{abstract}
  We present a systematic study of corrections to scaling in the
  self-organized critical forest-fire model. The analysis of the
  steady-state condition for the density of trees allows us to
  pinpoint the presence of these corrections, which take the form of
  subdominant exponents modifying the standard finite-size scaling
  form.  Applying an extended version of the moment analysis
  technique, we find the scaling region of the model and compute
  the first non-trivial corrections to scaling.
\end{abstract}

}

\section{Introduction}

The term self-organized criticality (SOC) \cite{jenssen98} refers to a
set of driven dissipative systems that, under the action of a very
small external driving, evolve into a critical state characterized by
avalanches broadly distributed in space and time, which lead to
divergent (power-law) response functions. Since its introduction by
Bak, Tang, and Wiesenfeld \cite{btw1}, the concept of SOC has been the
object of a very intense research activity, covering both theory and
numerical simulations.

Among the many models proposed so far exhibiting SOC behavior, the
forest-fire model (FFM) \cite{bak92,drossel92,johansen94,clar94} is
one of the most simply defined and well understood.  The FFM is a
three states cellular automaton defined on a $d$-dimensional
hypercubic lattice. Each site of the lattice is occupied either by a
tree, by a burning tree, or is empty.  Every time step, the cellular
automaton evolves according to the following set of rules: (i) each
burning tree becomes an empty site; (ii) every tree with at least one
burning nearest neighbor becomes a burning tree; (iii) a tree becomes
a burning tree with probability $f$, irrespective of its nearest
neighbors; (iv) an empty site becomes a tree with probability $p$.
The FFM possesses two characteristic time scales \cite{drossel92}: the
average time for a tree to grow $1/p$ and the average time between
fires $1/f$. In the limit of the double infinite time scale
separation, $1\gg p\gg f$, the model displays critical behavior
\cite{drossel92,clar94}: i.e., fires are distributed according to
power laws. The magnitudes characterizing a fire are the total number
of trees burnt $s$, and the total time duration of the fire $t$
(measured as the total number of parallel updatings of the algorithm).
In the critical state, with $p/f\gg1$, the probability distributions of
sizes and times have been observed to follow the standard finite-size
scaling (FSS) hypothesis \cite{cardy88}:
\begin{eqnarray}
  P(s,\theta) &=& s^{-\tau_s} 
{\cal F} \left(\frac{s}{\theta^{\lambda_s}}\right),
  \label{eq:fss}\\ 
  P(t,\theta) &=& t^{-\tau_t} 
{\cal G} \left(\frac{t}{\theta^{\lambda_t}}\right),
  \label{eq:fsst} 
\end{eqnarray}
where $\theta=p/f$ is the critical parameter of the model
\cite{grassberger93}, and $\tau_x$ and $\lambda_x$ are scaling exponents
characterizing the critical state \cite{noteff}. Finally, ${\cal
  F}(z)$ and ${\cal G}(z)$ are cut-off functions that are constant for
$z\to0$ and decay exponentially fast for $z\to\infty$.

The precise determination of critical exponents is a relevant issue in
order to firmly establish universality classes and the upper critical
dimension, that on their turn are fundamental in the theoretical
understanding of the critical nature of the model.  While the
numerical determination of the overall power law behavior is a
relatively easy task, a very accurate determination of critical
exponents from numerical simulations can suffer from strong systematic
biases due to the distribution's lower and upper cut-offs.  More
subtly, the assumption of the FSS form does not take into account the
presence of corrections to scaling due to subdominant exponents. These
corrections are more evident for small values of the various
magnitudes and for deviations from pure criticality ($\theta^{-1}\neq 0)$.  On
the other hand, for a sufficiently large value of $\theta$, one can safely
assume that the scaling \equ{eq:fss}-\equ{eq:fsst} is essentially
correct. Let us then define the {\em scaling regime} of the model by
the parameter $\theta_{\rm scal}$, defined such that the single scaling
picture is correct for $\theta>\theta_{\rm scal}$; in principle, $\theta_{\rm scal}$
is a magnitude which depends on the microscopic details of the model.
However, the value of $\theta_{\rm scal}$ is in general unknown, and when
analyzing numerical data, there is no {\em a priori} way to ascertain
whether the range of $\theta$ at our disposal is large enough.

In this paper, we will show that in the stationary state of the forest
fire model, the presence of scaling corrections arises naturally.  The
analytical inspection of the steady state condition points out the
presence of subdominant scaling corrections and calls for an extended
scaling framework allowing the evaluation of the scaling regime and
the various corrections to scaling present in the model.  The proper
treatment of scaling corrections permits a more precise estimate of
the leading exponents.  In order to analyze the occurrence of
correction to scaling in a systematic way, we generalize the powerful
moment analysis introduced in Refs.~\cite{men,tebaldi99} to a more
general scaling form.  Within this framework, we are able to
estimate the value $\theta_{\rm scal}$, above which the simple form
\equ{eq:fss} is meaningful. We thus obtain corrected exponents, and
the values of the first subdominant exponents.

The paper is organized as follows: In Section II, by analyzing the
steady state condition, we show the ineluctable emergence of
subdominant corrections to scaling in the FFM. In Sec. III, we review
the moment analysis technique, and outline its extension to
probability distributions with subdominant terms. Sec. IV provides
numerical evidence of our results by means of extensive simulations of
the FFM in $d=2$. Finally, our conclusions are summarized in Sec. V.

\section{Stationarity condition and scaling corrections}

The necessity to include corrections to scaling indeed arises
naturally in the FFM, by just considering the steady-state condition
of the model \cite{drossel92,clar94}. For any value of $\theta$, at large
times the FFM sets in a steady state characterized by an average
constant density of trees, $\rho_t$, and empty sites, $1-\rho_t$ (the
density difference after and before a fire is negligible, being of
order $\theta/L^2$). The density of trees is known to display the
asymptotic behavior at large $\theta$ \cite{clar94,grassberger93}
\begin{equation}
  \rho_t=\rho_t^\infty  - a \theta^{-\alpha}.
  \label{eq:density}
\end{equation}
Computer simulations in $d=2$ provide the values $\rho_t^\infty \simeq0.408$ and
$\alpha\simeq0.5$ \cite{grassberger93}. In the steady-state, and for a fixed $\theta$
value, the average number of growing trees, $p (1-\rho_t ) L^d$, must
equal the average number of burnt trees, $f \rho_t \left< s\right>_\theta
L^d$, where $\left< s \right>_\theta$ is the average size of a fire.
Therefore, the mean number of trees burnt by a fire is given by
\begin{equation}   
  \left< s \right>_\theta  = \theta \frac{1-\rho_t}{\rho_t}.  
  \label{eq:aversizesimple} 
\end{equation} 
In the limit $\theta\to\infty$, the mean tree density reaches its critical value
$\rho_t^\infty $, and we recover the usual relation $\left< s \right>_\theta \sim \theta = p/f$
\cite{drossel92}. However, for any finite value of $\theta$, the system is
subcritical and $\rho_t<\rho_t^\infty $. Substituting the expression
\equ{eq:density} into \equ{eq:aversizesimple}, we obtain 
\begin{eqnarray}
  \left< s \right>_\theta  &=& \theta \times \frac{1- \rho_t^\infty +
    a \theta^{-\alpha}}{\rho_t^\infty  - a \theta^{-\alpha}} \nonumber
    \\ 
    &\simeq&  \frac{1- \rho_t^\infty }{\rho_t^\infty } \theta +
  \frac{a}{(\rho_t^\infty)^2}  \theta^{1-\alpha} + {\cal
    O}(\theta^{1-2\alpha}). 
  \label{eq:aversizecomplex}
\end{eqnarray}
That is, neglecting corrections of order
$\theta^{1-2\alpha}$ (which is valid since $\alpha\sim0.5$), the form
of the average fire size for finite $\theta$ is
\begin{equation}
  \left< s \right>_\theta = C_1 \theta + C_2 \theta^{1-\alpha}, 
  \label{eq:moment1}
\end{equation} 
where the $C_k$ are constants independent of $\theta$.  Inspection of
Eq.~\equ{eq:moment1} proves that it is impossible to obtain such a $\theta$
dependence for the average avalanche size with an FSS of the form
\equ{eq:fss}. We are therefore forced to admit a more complex scaling
form.  These corrections to scaling, which on the other hand are
well-known in the field of equilibrium and non-equilibrium critical
phenomena \cite{cardy96}, take the form of subdominant corrections to
the leading (infinite $\theta$) scaling form of the probability
distributions. The most general form of these corrections is
\begin{equation}
  P(x, \theta) = x^{-\tau_x} {\cal F}  \left(\frac{x}{\theta^{\lambda_x}}\right) + x^{-\tau_x^*} 
  {\cal F}^* \left(\frac{x}{\theta^{\lambda_x^*}}\right) + \ldots .
  \label{eq:correction}
\end{equation}
In the last equation, $\tau_x^*$ and $\lambda_x^*$ are subdominant exponents,
correcting the infinite $\theta$ behavior, and ${\cal F}^*(z)$ is a cut-off
function that decays faster than ${\cal F}(z)$ when $z\to\infty$. In this
way, and for fixed $\theta$, the effects of the corrections are expected to
be more noticeable for small values of $x$.  The ellipsis denotes
other possible corrections, which are of lower order compared to the
first one. 

In this perspective, a very accurate measurement of critical exponents
cannot escape the precise knowledge of the extent of the intermediate
region in which scaling corrections are still noticeable. In
particular, a method of analysis which takes into account the presence
of subdominant exponents is required for a fully consistent analysis
of the scaling properties at finite values of $\theta$.

\section{Moment analysis}

The determination of the scaling exponents for the FFM has been
performed most often in previous works by a direct measurement of the
slope of a log-log plot
\cite{drossel92,clar94,grassberger93,clar96,christensen93}. This
procedure yields the exponent $\tau_s$ by means of a straightforward
linear regression. The exponent $\lambda_s$ is then computed by imposing the
constraint \equ{eq:fss} for different values of $\theta$, using the
previously computed value of $\tau_s$ \cite{clar94}.

Even though with this procedure (sometimes supplemented with
extrapolations and/or local slope analysis) one can determine the
exponents within a $10\%$ accuracy, its performance is affected by the
existence of the upper and lower cutoffs, which render difficult its
application. Moreover, any binning performed to smooth the numerical
distributions can lead to biases very difficult to assess. In this
respect, it is better to use analysis techniques that use the whole
set of data (not only the power law regime) and contain explicitly the
system-size dependency. In the SOC field, the moment analysis has been
introduced by De Menech {\em et al.} in the context of the two
dimensional Bak-Tang-Wiesenfeld model \cite{men,tebaldi99} and has
been successfully applied to both deterministic and stochastic models
\cite{chessa99,granada,pv99jpa,lubeck99}.  In the following we
introduce the moment analysis and extend the method in order to deal
with scaling forms that make explicit the presence of subdominant
corrections.

\subsection{Single scaling form}

In this section, we concentrate in the moment analysis of the fire
size, following Refs.~\cite{men,tebaldi99}. We start with a
distribution fulfilling the scaling form \equ{eq:fss}.  The $q$-th
moment of the distribution is defined by $\left<s^q\right>_\theta = \int_1^\infty
s^q P(s,\theta) ds$. Inserting the scaling form of $P$ into this expression
yields the $\theta$ dependence
\begin{equation}
  \left<s^q\right>_\theta = \theta^{\lambda_s(q+1-\tau_s)} \int_{\theta^{-\lambda_s}}^\infty  z^{q-\tau_s}
  {\cal F}(z) dz,  
  \label{eq:scalsimple}
\end{equation}
where we have used the transformation $z=s/ \theta^{\lambda_s}$. For large values
of $\theta$, and provided that $q >\tau_s-1$, the lower limit of the integral
in \equ{eq:scalsimple} can be replaced by $0$. We then have
$\left<s^q\right>_\theta \sim \theta^{\lambda_s(q+1-\tau_s)}$. In general, we can write
$\left<s^q\right>_\theta \sim \theta^{\sigma_s(q)}$, where the exponents $\sigma_s(q)$ can be
obtained as the slope of a log-log plot of $\left<s^q\right>_\theta$ as a
function of $\theta$.  Comparing with \equ{eq:scalsimple}, one has $\langle
s^{q+1}\rangle_\theta /\langle s^q\rangle_\theta   \sim \theta^{\lambda_s}$ or $\sigma_s(q+1)-\sigma_s(q)=\lambda_s$, so that the
slope of $\sigma_s(q)$ as a function of $q$ is the cutoff exponent; i.e.
$\lambda_s=\partial\sigma_s(q)/\partial q$.  This is not true for small $q$, because the
integral in \equ{eq:scalsimple} is dominated by its lower cutoff.  In
particular, the lower cutoff becomes important for $q \leq\tau_s-1$.  Once
the exponent $\lambda_s$ is known, we can estimate $\tau_s$ from the scaling
relationship $(2-\tau_s)\lambda_s=\sigma_s(1)$.

The results of the moment analysis must finally be checked by means of
a data collapse analysis. The initially assumed FSS hypothesis
\equ{eq:fss} has to be verified, and must be consistent with the
calculated exponents. This can be done by rescaling $s\to s/ \theta^{\lambda_s}$
and $P(s, \theta)\to P(s, \theta)\theta^{\lambda_s \tau_s}$. Data for different values of $\theta$
must then collapse onto the same universal curve if the FSS hypothesis
is to be satisfied. Complete consistency between the methods gives the
best collapse with the exponents obtained by the moment analysis.

\subsection{Moment analysis with corrections to scaling}

Let us now develop the formalism of the moment analysis for a
distribution with corrections to scaling of the form
\equ{eq:correction}, where we will only keep the first non-trivial
correction. By plugging this form into the definition of the $q$-th
moment, we obtain
\begin{eqnarray}
  \left< s^q \right>_\theta &=& \int_1^\infty \!\! s^{-\tau_s+q} {\cal
  F}  \left(\frac{s}{\theta^{\lambda_s}}\right) d s +  \int_1^\infty
  \!\!  s^{-\tau_s^* +q} {\cal F}^* 
  \left(\frac{s}{\theta^{\lambda_s^*}}\right) d s \nonumber \\
  &=& \theta^{\lambda_s(q+1-\tau_s)}
  \int_{\theta^{-\lambda_s}}^\infty z^{-\tau_s+q} {\cal F}(z) d z
  \nonumber \\
  &+&  \theta^{\lambda_s^*(q+1-\tau_s^*)}
  \int_{\theta^{-\lambda_s^*}}^\infty z^{-\tau_s^*+q} {\cal 
    F}^*(z) dz.
\end{eqnarray}
In the integrals of the previous expression we have explicitly
written the dependence on the lower cut-off. For $\theta$ sufficiently
large and $q>\max(\tau_s, \tau_s^*)-1$, the lower limits tend to zero,
and thus we expect the integrals to be independent of $\theta$.
However, we cannot discard in general a possible dependence on $q$
(through the exponent in the integrand). We have therefore
\begin{equation}
  \left< s^q \right>_\theta = C(q)  \theta^{\lambda_s(q+1-\tau_s)} + C^*(q) \theta^{\lambda_s^*(q+1-\tau_s^*)},
  \label{eq:momenttotal}
\end{equation}
where we have defined the constants (independent of $\theta$)
\begin{eqnarray*}
  C(q) &=& \int_0^\infty z^{-\tau_s+q} {\cal F}(z) d z, \\
  C^*(q) &=& \int_0^\infty z^{-\tau_s^*+q} {\cal F}^*(z) dz.
\end{eqnarray*}

Analysis of the general Equation \equ{eq:momenttotal} is extremely
difficult, due to the impossibility to separate the two leading
behaviors $\theta^{\lambda_sq}$ and $\theta^{\lambda_s^*q}$. In order to achieve further
progress, we must somehow simplify relation \equ{eq:momenttotal}. To
do so, we proceed to make an ansatz, whose validity will have to be
numerically verified a posteriori. The ansatz consists in assuming the
identity
\begin{equation}
  \lambda_s=\lambda_s^*,
\end{equation}
that is, the cut-off exponents do not suffer from corrections.  The
physical interpretation of this single cut-off exponent for both the
leading and subdominant terms in the size probability distribution is
related to the presence of a unique and well-defined divergent
characteristic size in the avalanche evolution.  Under this
assumption, Eq.~\equ{eq:momenttotal} becomes
\begin{equation}
  \left< s^q \right>_\theta = \theta^{\lambda_s(q-1)} \left[C(q)  \theta^{\lambda_s(2-\tau_s)} + C^*(q)
    \theta^{\lambda_s(2-\tau_s^*)} \right].
  \label{eq:momentansatz}
\end{equation}
Specializing this relation to $q=1$ we obtain
\begin{equation}
  \left< s \right>_\theta = \left[C(1) \theta^{\lambda_s(2-\tau_s)} + C^*(1)
    \theta^{\lambda_s(2-\tau_s^*)} \right].
\end{equation}
Comparing now with the expression for the average fire size (first
moment), Eq.~\equ{eq:moment1}, we can identify $\lambda_s(2-\tau_s)=1$ and
$\lambda_s(2-\tau_s^*)=1-\alpha$, from which we obtain the exponents
\begin{equation}
  \tau_s=2-1/ \lambda_s, \qquad \tau_s^*=2-(1-\alpha)/ \lambda_s.
  \label{eq:taus}
\end{equation}
Using the previous relations, we can express Eq.~\equ{eq:momentansatz} 
as a function of the exponents $\alpha$ and $\lambda_s$ alone:
\begin{equation}
  \left< s^q \right>_\theta = \theta^{\lambda_s(q-1)+1} \left[C(q) + C^*(q) \theta^{-\alpha} \right].
  \label{eq:momentfinal}
\end{equation}

Eq.~\equ{eq:momentfinal} suggests the correct strategy to work out the
moment analysis. Firstly, we observe that the quantity 
\begin{equation}
  \Gamma_q(\theta)\equiv \frac{\left< s^q \right>_\theta}{\theta^{\lambda_s(q-1)+1}} = C(q) + C^*(q)
  \theta^{-\alpha} 
  \label{eq:gammaq}
\end{equation}
depends only on $\theta^{-\alpha}$. We can use this fact to verify the validity
of the ansatz $\lambda_s^*=\lambda_s$ by plotting $\Gamma_q(\theta)$ as a function of
$\theta^{-\alpha}$ for different values of $q$, and checking whether or not the
plots have linear dependence. Secondly, we note that the second term
between brackets in the r.h.s. of \equ{eq:momentfinal} decreases with
increasing $\theta$. For $\theta$ sufficiently large, this second term is
negligible with respect to the constant $C(q)$, $\left< s^q \right>_\theta$
has a pure power-law dependence and we can proceed to compute $\lambda_s$ by
means of linear regressions. Indeed, this observation allows us to
define quantitatively the scaling region of the model: assuming that
the ratio $C(q)/C^*(q)$ does not depend strongly on $q$, we define
$\theta_{\rm scal}$ as the value of the scaling parameter for which
\begin{equation}
  \frac{C^*(1)}{C(1)} \theta_{\rm scal}^{-\alpha} \leq r,
  \label{eq:criterion}
\end{equation}
with $r$ some (arbitrary) small number. For $\theta>\theta_{\rm scal}$, the
approximation $\left< s^q \right>_\theta \simeq \theta^{\lambda_s(q-1)+1} C(q)$, and
therefore the single scaling form \equ{eq:fss} is correct, within a
precision of order $r$. One can thus proceed to compute the quantity
$\sigma_s(q)$ by means of regressions limited to values of $\theta>\theta_{\rm
  scal}$, determine $\lambda_s$ by differentiation and, using
Eq.~\equ{eq:taus}, estimate the rest of the exponents.

\begin{figure}[t]
  \centerline{\epsfig{file=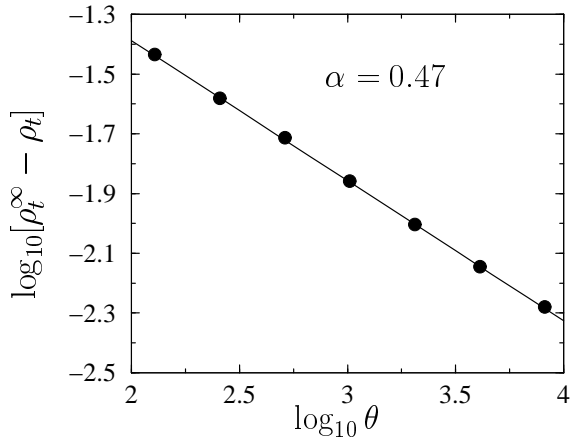, width=7.5cm}}
  \vspace*{0.5cm}
  \caption{Scaling of the average density of trees as a function of
    the parameter $\theta=p/f$.}
  \label{fig:treedensity}
\end{figure}

\section{Numerical Results}

In order to check numerically our arguments, we have performed
extensive numerical simulations of the FFM in $d=2$, using the
algorithm described in Ref.~\cite{grassberger93}. Starting from an
arbitrary initial configuration, we update the lattice according to
the following rules: (i) select at random a site in the lattice; if
the site contains a tree, burn it and all the trees that belong to its
same forest cluster; if the site is empty, proceed to step ii; (ii)
select at random $\theta$ sites; if a site is empty, grow a tree on it; if
it contains a tree, do nothing. It is easy to see that these set of
rules are equivalent to the original definition of the FFM, in the
limit $p=0^+$ and finite $p/f=\theta$. For large $\theta$, we thus ensure the
double infinite time scale separation condition.  The system sizes
considered are up to $L=19000$ and the values of $\theta$ range from $128$
to $32768$.  Results are averaged over $10^7$ nonzero fires.

\begin{figure}[t]
  \centerline{\epsfig{file=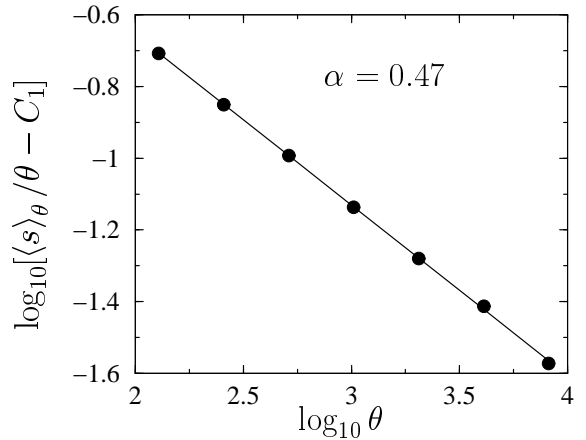, width=7.5cm}}
  \vspace*{0.5cm}
  \caption{Scaling of the average fire size as a function of
    the parameter $\theta=p/f$.}
  \label{fig:firstmoment}
\end{figure}

\subsection{Average density of trees}

In the first place, we study the average density of trees as a
function of the parameter $\theta$. After discarding a sufficiently large
number of fires (usually $5\times10^5$) to ensure that the system is in a
steady-state, we compute the average number of trees, per unit area,
left {\em after} each fire. The measured $\rho_t$ is fitted to the
functional form $\rho_t^\infty - a \theta^{-\alpha}$ using the Levenberg-Marquardt
non-linear fitting algorithm \cite{recipes}. We obtain a critical
asymptotic density of trees $\rho_t^\infty =0.4084\pm0.0005$, and an exponent
$\alpha=0.47\pm0.01$, in good agreement with previous results \cite{clar94}.
In Figure~\ref{fig:treedensity} we have checked the asymptotic form of
the average tree density by plotting $\log_{10} (\rho_t^\infty -\rho_t)$ as a
function of $\log_{10} \theta$.

In Fig.~\ref{fig:firstmoment} we check the validity of
Eq.~\equ{eq:moment1}. The parameters computed, using again a
non-linear curve fitting, are $C_1\simeq0.854$, $C_2\simeq1.973$, and
$\alpha=0.47\pm0.01$.  Again, we observe a very good fit to the predicted
form. In view of this results, we select the value $\alpha=0.47$ for the
computations to follow.

\subsection{Size probability distribution}

\begin{figure}[t]
  \centerline{\epsfig{file=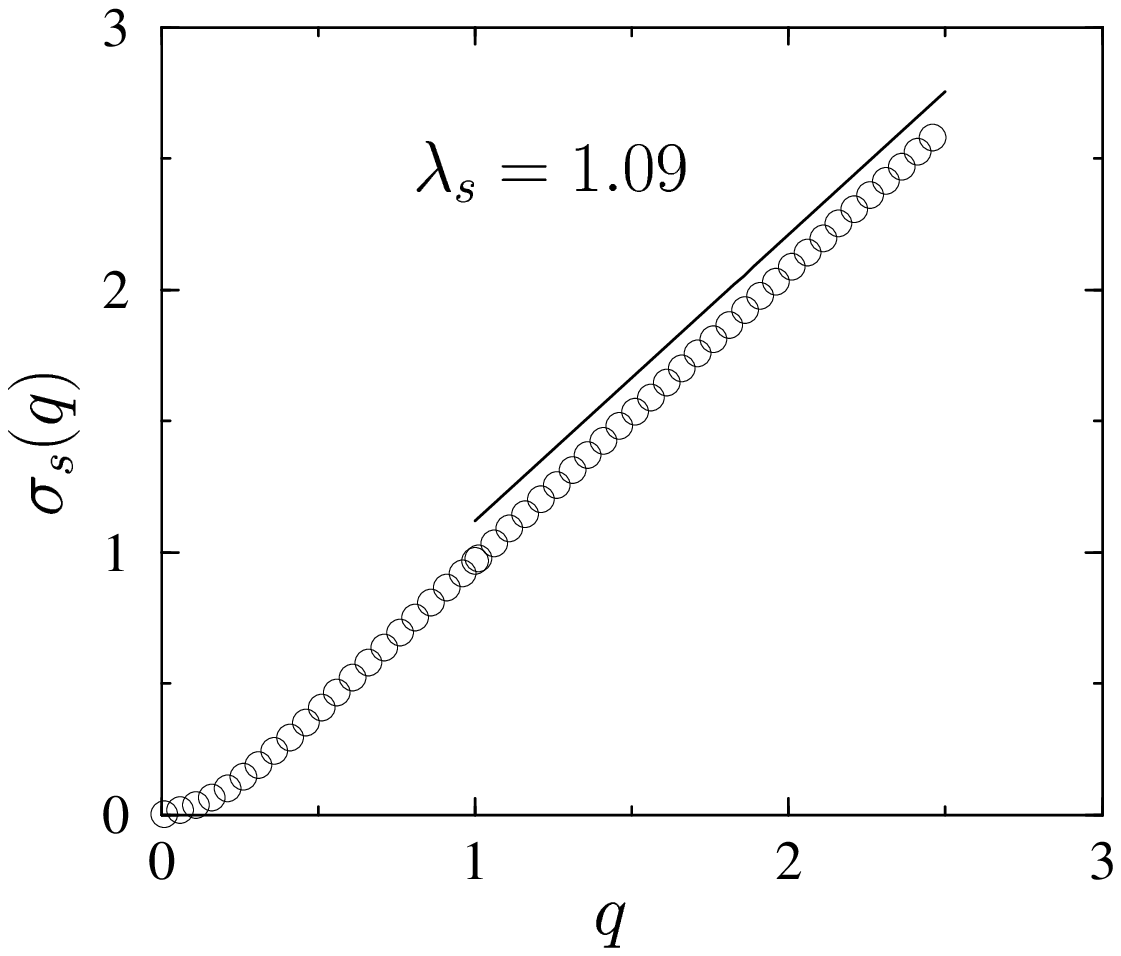, width=7.5cm}}
  \vspace*{0.5cm}
  \caption{Plot of $\sigma_s(q)$, computed from linear regressions from
    $\theta=4096$ to $32768$. The slope yields the exponent
    $\lambda_s=1.09\pm0.01$.} 
  \label{fig:sigmas}
\end{figure}

Once we have verified the likelihood of corrections to scaling in the
first moment of the fire size distribution, we proceed to analyze the
size probability distribution.  The first step it to compute the
threshold $\theta_{\rm scal}$ using the criterion \equ{eq:criterion}. We
arbitrarily fix the parameter $r=0.05$; for this value, together with
the estimates of $C_1=C(1)$ and $C_2=C^*(1)$ obtained by analyzing
$\left< s \right>_\theta$, we estimate $\theta_{\rm scal} \geq 3000$. For values of
$\theta$ larger than $3000$; therefore, the single FSS form \equ{eq:fss}
can be assumed to be valid, and we can proceed along the standard
moment analysis technique. In Fig.~\ref{fig:sigmas} we plot the
moments $\sigma_s(q)$ computed from linear regressions of $\log_{10}
\left<s^q\right>_\theta$ as a function of $\log_{10} \theta$, for values of $\theta$
between $4096$ and $32768$. The slope of this plot yields the exponent
$\lambda_s=1.09\pm0.01$; finally using the relations \equ{eq:taus} with
$\alpha=0.47$, we obtain $\tau_s=1.08\pm0.01$ and $\tau_s^*=1.51\pm0.02$. A summary
of results is presented in Table~\ref{table}.

\begin{figure}[t]
  \centerline{\epsfig{file=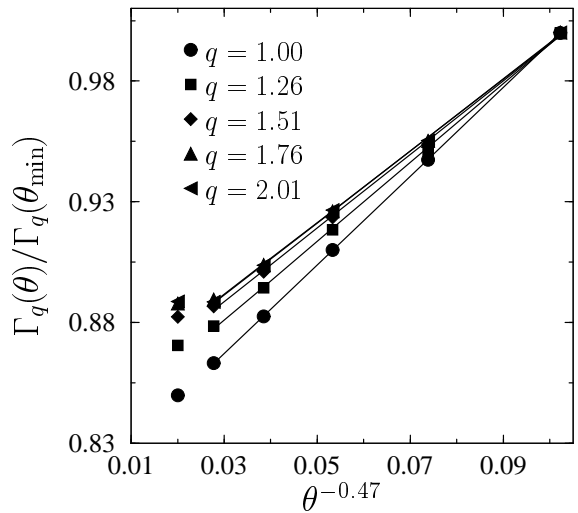, width=7.5cm}}
  \vspace*{0.5cm}
  \caption{Rescaled $q$-th moment
    $\Gamma_q(\theta)/\Gamma_q(\theta_{\rm min})$ as 
    function of $\theta^{-0.47}$. The good linear fit for small
    $\theta$ validates the ansatz $\lambda_s^*=\lambda_s$. The full
    lines are guides to the eye.}
  \label{fig:gammas}
\end{figure}

\begin{table}[b]
\begin{tabular}{p{0.5in}ccccc}

 & $\tau_s$   & $\lambda_s=\lambda_s^*$   & $\tau_s^*$   & $\tau_t$           & $\lambda_t$ \\
\hline
Slope\\Analysis
& $1.14(3)$ & $1.15(3)$ & ---     &$1.27(7)$ & $0.58$\\  
Moment \\Analysis
& $1.08(1)$ & $1.09(1)$ & 1.51(2) &$1.27(1)$ & $0.59(1)$\\  
\end{tabular}
\caption{Critical exponents for the FFM model, obtained through the
  slope analysis, Ref.~\protect\cite{clar94}, and by means of the moment
  analysis. Figures in parenthesis denote statistical uncertainties.} 
\label{table}
\end{table}

Once we have computed the exponent $\lambda_s$, we can check {\em a
  posteriori} the validity of the ansatz $\lambda_s^*=\lambda_s$. We
do so by plotting the quantity $\Gamma_q(\theta)/\Gamma_q(\theta_{\rm
  min})\sim\left< s^q \right>_\theta / \theta^{\lambda_s(q-1)+1}$ as a
function of $\theta^{-\alpha}$, with $\alpha=0.47$, for several values
of $q$, Fig.~\ref{fig:gammas}. For large values of $\theta^{-\alpha}$,
we observe a very good linear relationship.  The goodness of the fit
decreases for large $q$ and large $\theta$ (small $\theta^{-\alpha}$)
because in both cases, the $q$-th moment is dominated by the largest
avalanches, of which there is poorer statistics.  We conclude
therefore that indeed the assumption $\lambda_s^*=\lambda_s$ is indeed
well justified for the FFM.

\begin{figure}[t]
  \centerline{\epsfig{file=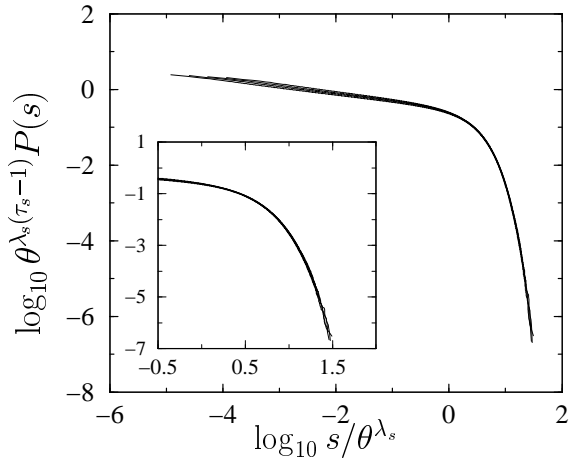, width=7.5cm}}
  \vspace*{0.5cm}
  \caption{Data collapse analysis of the integrated fire size
    distribution. $\theta=4096, 8192, 16384,$ and $32768$. Exponents used:
    $\lambda_s=1.09$, $\tau_s=1.08$. Inset: detail of the tail of the distribution.}
  \label{fig:collapse}
\end{figure}

The presence of corrections of the form \equ{eq:correction} make
impossible to use a standard data collapse to inspect the accuracy of
our results in the whole range of $\theta$ values. However, for $\theta>\theta_{\rm
  scal}$, is reasonable to expect a good collapse to the single form
\equ{eq:fss}. We have plotted this data collapse in
Figure~\ref{fig:collapse}, for the integrated size distributions. The
collapse for the exponential tail of the distribution is quite
remarkable. On the other hand, it is poorer for small values of $s$.
This effect is due to the very presence of corrections to scaling,
whose influence is stronger for small $s$.

Our method provides values which correct previous estimates (namely,
in our notation, $\tau_s=1.14\pm0.03$ and $\lambda_s=1.15\pm0.03$,
Ref.~\cite{clar94}) by a $5\%$.  As explained before, the discrepancy
is due to the fact that in Ref.~\cite{clar94} exponents were computed
by directly measuring the slope of the probability distributions, a
method which is usually less accurate.  Our value of $\lambda_s$, on the
other hand, agrees better with the result reported in
Ref.~\cite{grassberger93}, which was obtained by a method closer in
spirit to the moment analysis.

\subsection{Time probability distribution}

To complete the study of the FFM, we proceed in this section to apply
the moment analysis to the fire time distribution. Here there is no
{\em a priori} clue about the possible existence of corrections to
scaling. We will therefore assume the simple FSS form \equ{eq:fsst}
and perform the analysis for values of $\theta$ larger than $\theta_{\rm scal}$.

\begin{figure}[t]
  \centerline{\epsfig{file=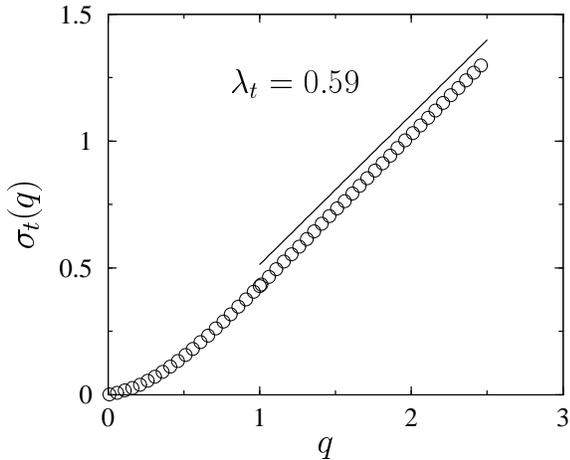, width=7.5cm}}
  \vspace*{0.5cm}
  \caption{Plot of $\sigma_t(q)$, computed from linear regressions from
    $\theta=4096$ to $32768$. The slope yields the exponent
    $\lambda_t=0.59\pm0.01$.} 
  \label{fig:sigmast}
\end{figure}

\begin{figure}[t]
  \centerline{\epsfig{file=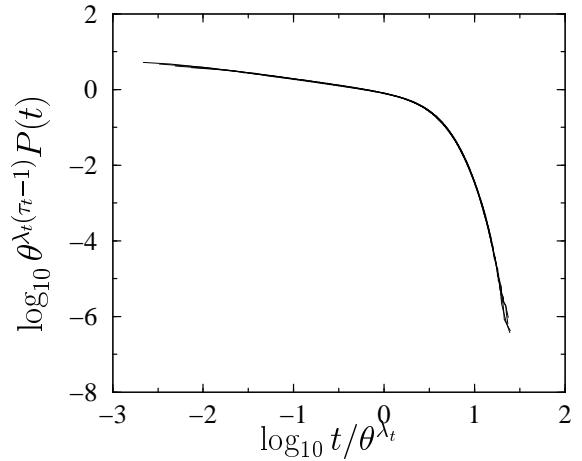, width=7.5cm}}
  \vspace*{0.5cm}
  \caption{Data collapse analysis of the integrated fire time
    distribution. $\theta=4096, 8192, 16384,$ and $32768$. Exponents used:
    $\lambda_t=0.59$, $\tau_t=1.27$.}
  \label{fig:collapsetime}
\end{figure}

Along the same lines followed for the size distribution, we define the
$q$-th time moment $\left<t^q\right>_\theta = \int_1^\infty t^q P(t,\theta) dt$. In this
case, we have $\left<t^q\right>_\theta \sim \theta^{\sigma_t(q)}$, with $\lambda_t=\partial\sigma_t(q)/\partial
q$ and $\tau_t$ given by the relation $(2-\tau_t)\lambda_t=\sigma_t(1)$. In
Figure~\ref{fig:sigmast} we plot $\sigma_t(q)$ as a function of $q$,
computed by linear regression for the largest values of $\theta$. From the
slope of this plot we obtain $\lambda_t=0.59\pm0.01$, and using this value on
the precedent scaling relation, we obtain $\tau_t=1.27\pm0.01$. The data
collapse with these exponents of the integrated time distribution is
shown in Figure~\ref{fig:collapsetime}. In this case, and on the
contrary to the size distribution, the collapse is perfect for all
values of $t$, which proves the irrelevance of corrections to scaling
in the distribution of this magnitude.

It is interesting to note that our results match quite closely the
results in \cite{clar94}, namely $\lambda_t=0.58$ and $\tau_t=1.27\pm0.07$. This
fact is accounted for by the method employed in \cite{clar94} to
compute $\lambda_t$, that is, an analysis of the lifetime of the largest
fire, $T_{\rm max}$, as a function of $\theta$. This procedure indeed
amounts to an estimation of the cut-off exponent of the time
distribution, and is presumably less error-prone that a direct
measurement of the initial slope of the distribution.

\section{Conclusions}

In summary, in this paper we have shown that subdominant scaling
corrections are inescapable in the forest-fire model.  The analytical
analysis of the stationarity condition shows that scaling corrections
to a simple FSS form of the fires distribution must be included in
order to account for the model behavior at finite values of $\theta$. In
this perspective, we have proposed a  method to explore
corrections to the finite-size scaling hypothesis in the forest-fire
model. The method, based in an extension of the moment analysis,
allows in principle the determination of the scaling regime of the
models, as well as the computation of the first order corrections to
scaling.  Applying our method, we have been able to compute
numerically corrected values to the scaling exponents, summarized in
Table~\ref{table}, and estimate the first nontrivial corrections.  We
note that our approach is complementary with previous studies of
deviation from scaling due to finite-size effects (small $L$ compared
with $\theta$) \cite{schenk99}.

As a final remark, it is interesting to point out that the present
method can also be applied to standard sandpile models, defined on a
lattice of size $L$ with open boundary conditions.  In this case,
however, the applicability of the method is hindered by the
availability of a smaller range of values of the scaling parameter
$L$. Interestingly, preliminar results with medium system sizes
indicate that the ansatz $\lambda_s=\lambda_s^*$ may be violated in sandpiles.
This fact can be related to the more complex structure of avalanches
in sandpiles (compared with the percolation-like fires in the FFM),
that induce the presence of more than one characteristic avalanche
size. Unfortunately, the violation of the ansatz renders the
computation of the corrections considerably much harder. Work is
underways to explore the full structure of the corrections to scaling
in sadpiles.

\section*{Acknowledgements}

We thank A. Stella for helpful discussions.  This work has been
supported by the European Network under Contract No.  ERBFMRXCT980183.


\begin{references}
\bibitem{noteff}
We have chosen a notation more conventional in sandpile models. The equivalence
  with the notation in Refs.~\cite{drossel92,clar94,grassberger93}, based on
  percolation theory, is $\tau_s\to\tau-1$, $\lambda_s\to\lambda$, $\tau_t\to
  b$, and $\lambda_t\to\nu'$.
\end{references}
\end{document}